\documentstyle[graphicx,float]{article}

\begin{document}
\title{The two-particle two-slit experiment}
\date{}
\author{Pedro Sancho \\ Centro de L\'aseres Pulsados CLPU \\ Parque Cient\'{\i}fico, 37085 Villamayor, Salamanca, Spain}
\maketitle
\begin{abstract}
Identical two-particle interferometry provides a scenario where
interference and exchange effects manifest at once. We present a
detailed calculation of the detection patterns in the two-particle
two-slit experiment by extending Feynman's Gaussian slit approach to
multi-mode states. We show the existence of two regimes depending on
the mode distributions. In one of them we find a novel behavior for
bosons, with detection patterns almost equal to those of
distinguishable particles although the overlapping is large. As a
byproduct of our analysis we conclude that the interaction at the
diffraction grating is an efficient method to increase the initial
overlapping. We propose a scheme, within the reach of present day
technology, to experimentally verify our results.
\end{abstract}

\section{Introduction}

Multi-photon interference, specially two-photon one, has been
extensively studied in Quantum Optics after the seminal
Hong-Ou-Mandel work (HOM) \cite{HOM,Lou}. More recently, similar
ideas have been presented for identical massive particles
\cite{Sil,SaJP,SaPR}. Identical multi-particle interference is
interesting because two fundamental principles of quantum theory,
interference and (anti)symmetrization of the states of identical
particles, act simultaneously in the system. Interference and
exchange effects can be observed at once.

In previous proposals of massive multi-particle interferometry no detailed
evaluation of the detection patterns has been presented. Such an
evaluation is necessary for comparison with the results of any
potential experimental test. For instance, in \cite{SaJP} a
calculation of the two-particle one-slit patterns has been carried
out, but with an oversimplified state. It is necessary to consider
more realistic states. We do it here for the two-slit arrangement.

The two-slit experiment is the archetypical example of quantum
interference. It has been used on countless occasions to illustrate
the wave properties of quantum systems. However, the massive
two-particle counterpart of the arrangement (with a detailed
evaluation of the expected patterns) has not been discussed in the
literature. As it is well known, the exact evaluation of the wave
function of a particle passing through a slit demands the use of
Fresnel's functions, complicating the understanding of the problem. In order
to carry out the discussion in mathematical terms as simple as
possible we resort to Feynman's Gaussian slit approximation,
used in the path integrals formalism \cite{fey}.
Moreover, to make the evaluation more realistic, we shall consider
particles in multi-mode states instead of the single-mode one
presented in \cite{fey}.

In the above approximation we can determine the two-particle
detection patterns, which show the existence of two regimes. For
similar mode distributions of the two particles (in a sense
discussed quantitatively later) the patterns of distinguishable
particles and bosons are almost equal and indistinguishable from the
experimental point of view. In contrast, fermions patterns are
clearly different. When the mode distributions are not similar we
have another regime where the three types of patterns are easily
distinguishable.

As we shall see later, in the first regime we have in general a
large overlapping. Then we should expect an important contribution
of the exchange effects and, as in the HOM experiment, a clearly
differentiated behavior of bosons and distinguishable particles. The
unexpected behavior found in that regime can be explained by the
simultaneous contribution of the large final overlapping to the form
of the two-particle wave function and its normalization, two factors
that cancel each other out. As a byproduct of the analysis we shall
show that the interaction at the diffraction grating increases the
initial overlapping of the particles.

In the final part of the paper we propose a scheme for the
experimental verification of two-atom two-slit interferometry. The source
of a single particle diffraction arrangement is replaced by one of
(anti)symmetrized two-particle states using techniques of atom
lasers \cite{Ess}, making possible the comparison of the
two-particle detection patterns of bosons and fermions with these of
distinguishable particles.

\section{General expressions for two-particle systems}

Our arrangement is a standard two-slit interference device, but with
the source emitting pairs of particles. The particles can be
distinguishable or identical. After the slits we place detectors
that determine the joint patterns by counting arrivals on
coincidence. We denote the two slits by $A$ and $B$. The two
particles are in the states $\psi (x)$ and $\phi (y)$, with $x$ and
$y$ their coordinates. If we consider detections in a plane parallel
to the slits we can reduce the problem to an one-dimensional one, as
we do here. The first wave function can be expressed as
\begin{equation}
\psi (x) =N_{\psi} (\psi _A(x)+\psi _B(x))
\end{equation}
with $ N_{\psi}$ the normalization coefficient, deduced from $\int
dx |\psi (x)|^2=1$ . A similar expression holds for $\phi$.

If the two particles are distinguishable the state of the two-particle system is a product one:
\begin{equation}
\Phi (x,y) =\psi (x)\phi (y)
\label{eq:dos}
\end{equation}
Note that this state is already normalized, $\int \int dxdy |\Phi
(x,y)|^2=1$, because of the previous normalization of $\psi$ and $\phi$.

When the two particles are identical the joint state is
\begin{equation}
\Psi (x,y) =N_{\Psi} (\psi (x) \phi (y) \pm \psi (y) \phi (x))
\end{equation}
In the double sign expression $\pm$ the upper one refers to bosons
and the lower one to fermions. The normalization factor is
\begin{equation}
N_{\Psi}=\frac{1}{(2\pm 2|<\psi |\phi >|^2)^{1/2}} \label{eq:nor}
\end{equation}
where, for the matter of simplicity, we have used the ket
notation $<\psi |\phi >=\int dx \psi ^* (x)\phi (x)$.

From these expressions we can derive the probability densities
of joint detection. In the distinguishable case, in order to obtain a more
compact presentation (in the comparison with the identical case,
where there is an interchange of the coordinates $x$ and $y$) it is
natural to present the results in a more symmetric way. In Eq.
(\ref{eq:dos}) the particles in states $\psi$ and $\phi$ are labeled
by $x$ and $y$. Similarly, we could denote them by $y$ and $x$
($\psi (y)$ and $\phi (x)$). Then the symmetric presentation of the
probability density for distinguishable particles is
\begin{equation}
P_{dis}(x,y)=\frac{1}{2} |\Phi (x,y)|^2 + \frac{1}{2} |\Phi (y,x)|^2
\end{equation}
For identical particles the density probability is $P(x,y)=|\Psi (x,y)|^2$. After a simple calculation we have
\begin{equation}
P(x,y)=2|N_{\Psi}|^2P_{dis}(x,y) \pm 2 |N_{\Psi}|^2 \sum _{i_1,i_2,i_3,i_4}^{A,B} Re(\psi ^*_{i_1}(x)\phi _{i_2}^*(y) \psi _{i_3}(y)\phi _{i_4}(x))
\end{equation}
In the above sum the indexes $i_1,i_2, \cdots $ run over the
two alternatives available for each particle, $A$ and $B$.

The probability density is given by the sum of two terms, the direct and
exchange ones. Note that there are sixteen exchange terms associated
with the eight alternatives present in the problem (the first alternative is the
particle in state $\psi$ going through slit $A$ and the particle in
state $\phi$ also passing through $A$,...).

\section{Feynman's Gaussian slit approximation}

Our next step is to derive the wave functions of the particles after the
slits, for instance, $\psi _A$. The source is placed at point $x=0$
and the middle point of the slit (with a width $2b$) at $x_0$. The
particle is initially in a multi-mode state described by the
distribution \cite{Ada}
\begin{equation}
f(k)=\frac{(4\pi)^{1/4}}{\sigma ^{1/2}} \exp(-k^2/2\sigma ^2)
\end{equation}
This is a Gaussian function with mean value $k_0=0$. The wave function in the position space is given by
\begin{eqnarray}
\psi _A (x,t)=(2\pi)^{-1}\int dk f(k) \exp(i(kx-k^2\hbar t/2m))=\nonumber \\
C(t)\exp \left( \frac{ -\sigma ^2 x^2 + i\hbar \sigma ^4 x^2t/m}{\mu
(t)} \right)
\label{eq:err}
\end{eqnarray}
with
\begin{equation}
C(t)=\pi ^{-1/4} \left( \frac{1}{\sigma} + \frac{i\hbar \sigma t}{m}  \right)^{-1/2} ; \mu (t)= 2\left( 1+\frac{\hbar ^2 \sigma ^4 t^2}{m^2}  \right)
\end{equation}
Equation (\ref{eq:err}) differs from Eq. (17) in \cite{Ada} by the
absence of a factor $2$ in the second term of the numerator of the
exponential.

The particle evolves free up to the slit. After the slit the wave
function can be expressed using the path integral formalism
\cite{fey}
\begin{equation}
\psi _A(x,t)=\int _{slit} dx_s K(x,t;x_s,t_s)\psi _A(x_s,t_s)
\end{equation}
Where the subscript $s$ refers to the coordinates of the slit
($x_s$) and the time when the particle reaches it ($t_s$). The
integration is over all the extension of the slit (for the rest of
the paper, when no explicit range of integration is included we
refer to $(-\infty , \infty)$). $K$ denotes the kernel of the
problem which is well known to be \cite{fey}
\begin{equation}
K(x,t;x_s,t_s)=\left(  \frac{m}{2\pi i\hbar (t-t_s)} \right)^{1/2} \exp \left( \frac{im(x-x_s)^2}{2\hbar (t-t_s)} \right)
\end{equation}
As signalled in the Introduction the above integral leads to
Fresnel's functions, making impossible to deal analytically with the
problem. Thus, to present a more intuitive physical picture of the
problem we avoid, following Feynman, the use of the Fresnel
functions by introducing the Gaussian slit approximation. In this
approach the integration is extended to all the axis, but with a
weight function $\exp (-(x_s-x_0)^2/2b^2)$ that reduces the relevant
values of the integrand to the proximity of the slit. Explicitly, we
have
\begin{equation}
\psi _A(x,t)=\int dx_s e^{-(x_s-x_0)^2/2b^2} K(x,t;x_s,t_s)\psi _A(x_s,t_s)
\end{equation}
After a simple calculation we obtain
\begin{equation}
\psi _A(x,t)= {\cal C} e^{imx^2/2\hbar (t-t_s)} e^{-(\alpha -
i\beta
) x^2} e^{-(\delta +i\gamma)x}
\label{eq:fun}
\end{equation}
with
\begin{equation}
{\cal C}=C(t_s) \left( \frac{m}{2i\hbar (t-t_s)(D+iF)} \right)^{1/2} e^{-x_0^2/2b^2} e^{G^2(D-iF)/4(D^2+F^2)}
\end{equation}
\begin{equation}
\alpha = \frac{DH^2}{4(D^2+F^2)} ; \beta = \frac{FH^2}{4(D^2+F^2)} ; \gamma = \frac{DGH}{2(D^2+F^2)} ; \delta = \frac{GHF}{2(D^2+F^2)}
\end{equation}
and
\begin{equation}
D=\frac{1}{2b^2}+ \frac{\sigma^2}{\mu};
F=-\frac{\hbar \sigma ^4t_s}{m\mu}- \frac{m}{2\hbar (t-t_s)};
G=\frac{x_0}{b^2}; H=\frac{m}{\hbar (t-t_s)}
\end{equation}

\section{Two-particle probability densities}

The expressions for the other wave functions are similar to Eq. (\ref{eq:fun}):
\begin{equation}
\psi _B(x,t)= {\cal C} e^{imx^2/2\hbar (t-t_s)} e^{-(\alpha - i\beta
) x^2} e^{(\delta +i\gamma)x}
\end{equation}
\begin{equation}
\phi _A(x,t)= \overline{\cal C} e^{imx^2/2\hbar (t-t_s)}
e^{-(\overline{\alpha}- i\overline{\beta }) x^2}
e^{-(\overline{\delta} +i\overline{\gamma})x}
\end{equation}
and
\begin{equation}
\phi _B(x,t)= \overline{\cal C} e^{imx^2/2\hbar (t-t_s)} e^{-(\overline{\alpha }- i\overline{\beta } ) x^2} e^{(\overline{\delta} +i\overline{\gamma})x}
\end{equation}
All the coefficients of the particle in state $\phi$ are denoted by an
overline. The width of the mode distribution in this case is
$\overline{\sigma}$, giving rise to different $\overline{D}$ and
$\overline{F}$ and, consequently, different $\overline{\cal C}$,
$\overline{\alpha}$,... The different widths of the initial
multi-mode states is the parameter we use to control the initial
overlapping of the two particles. On the other hand, $H$ is equal in
all the cases and $G$ changes its sign for the slit $B$, leading to
the modifications $\gamma \rightarrow -\gamma$,...,
$\overline{\delta} \rightarrow -\overline{\delta}$. Finally, $\exp
(imx^2/2\hbar (t-t_s))$ is equal for the four terms and will cancel
in the calculation of probability densities.

Using these expressions we can derive the normalization coefficients
and the detection patterns. The normalization coefficients are
\begin{equation}
N_{\psi}=\frac{1}{|{\cal C}|} \left( \frac{\alpha}{2\pi}
\right)^{1/4} (e^{\delta ^2/2\alpha}+e^{-\gamma ^2/2\alpha })^{-1/2}
\end{equation}
and a similar expression for $N_{\phi}$ with obvious modifications.
$N_{\Psi}$ is given by Eq. (\ref{eq:nor}), where $|<\psi |\phi >|^2$
can be calculated from
\begin{eqnarray}
\frac{1}{{\cal C}^* \overline{\cal C}N_{\psi}N_{\phi}} \left(
\frac{(\alpha + \overline{\alpha })+i(\beta -\overline{\beta})}{\pi} \right)^{1/2} <\psi |\phi >= \nonumber \\
\exp \left( \frac{((\delta + \overline{\delta })-i(\gamma
-\overline{\gamma}) )^2}{4((\alpha + \overline{\alpha })+i(\beta
-\overline{\beta}))} \right) + \exp \left( \frac{((\delta -
\overline{\delta })-i(\gamma + \overline{\gamma}) )^2}{4((\alpha +
\overline{\alpha })+
i(\beta -\overline{\beta}))} \right) + \nonumber \\
\exp \left( \frac{((\overline{\delta} -\delta )+i(\gamma
+\overline{\gamma}) )^2}{4((\alpha + \overline{\alpha })+i(\beta
-\overline{\beta}))} \right) + \exp \left( \frac{(-(\delta +
\overline{\delta })+i(\gamma -\overline{\gamma}) )^2}{4((\alpha +
\overline{\alpha })+i(\beta -\overline{\beta}))} \right)
\end{eqnarray}
Finally, we evaluate the detection patterns. From the experimental
point of view they are measured by placing a detector at a fixed
position $x$ and by moving another detector along the $y$-axis. We
choose $x=0$ and obtain
\begin{eqnarray}
P_{dis}(0,y)/(|{\cal C}|^2|\overline{\cal C}|^2N_{\psi}^2 N_{\phi}^2)=
2e^{-2\alpha y^2}(e^{-2\delta y}+e^{2\delta y})+ \nonumber \\
2e^{-2\overline{\alpha} y^2}(e^{-2\overline{\delta} y}+e^{2\overline{\delta} y})+ 4 e^{-2\alpha y^2}\cos (2\gamma y) + 4 e^{-2\overline{\alpha} y^2}\cos (2\overline{\gamma} y)
\end{eqnarray}
and
\begin{eqnarray}
(e^{(\alpha +\overline{\alpha})y^2}/|{\cal C}|^2|\overline{\cal C}|^2N_{\psi}^2 N_{\phi}^2)\sum _{i_1,i_2,i_3,i_4}^{A,B}
Re(\psi ^*_{i_1}(0)\phi _{i_2}^*(y) \psi _{i_3}(y)\phi _{i_4}(0))= \nonumber  \\
4e^{-(\delta + \overline{\delta})y} \cos((\beta -\overline{\beta})y^2+ (\overline{\gamma}-\gamma )y) + 4e^{(\delta + \overline{\delta})y} \cos((\beta -\overline{\beta})y^2 - (\overline{\gamma}-\gamma )y) + \nonumber \\
4e^{(\delta - \overline{\delta})y} \cos((\beta
-\overline{\beta})y^2+ (\overline{\gamma}+\gamma )y) + 4e^{-(\delta
- \overline{\delta})y} \cos((\beta -\overline{\beta})y^2 -
(\overline{\gamma}+\gamma )y)
\end{eqnarray}

\section{Results}

Using the above equations we can graphically represent the behavior
of the joint detection probability densities. First of all, we
choose numerical values for the parameters of the problem. Typical
values of the slit size and separation between the slits are of the
order of $1 \mu m$. Usual separations between the source and
diffraction grating are close to half a meter, and velocities
perpendicular to the plane of the diffraction grating vary in a wide
range ($10^2$ to $10^7 ms^{-1}$). For instance, for atoms typical
values are around $3 \times 10^3 m s^{-1}$. Then we can take for the
time of fly $10^{-4}$ to $10^{-5}s$ and $\hbar t_s/m \approx 1\mu
m^2$ (similarly, for $\hbar (t-t_s)/m$). The dispersion of the
initial wavepacket ($\sigma \approx \Delta k$) in the axis parallel
to the diffraction grating can be expressed in terms of the
dispersion of the particle velocity in that direction, $\sigma
\approx m\Delta v /\hbar$. Taking into account the perpendicular
velocity and source-grating separation we must have $\Delta v
\approx 10^{-1}ms^{-1}$ (the probability of reaching the slits for
particles with much larger parallel velocities is negligible). This
corresponds to $\sigma \approx 1 \mu m ^{-1}$.

We choose for our representation the values $b=0.1 \mu m$,
$x_0=0.4\mu m$. $\hbar t_s/m = \hbar (t-t_s)/m =0.2 \mu m^2$,
$\sigma =1 \mu m^{-1}$ and $\overline{\sigma }=2 \mu m^{-1}$. The
most relevant fact (see Fig. 1) is that the curves for bosons and
distinguishable particles are almost equal. In contrast, the
behavior of fermions is clearly different.

Next, we represent (see Fig. 2) the detection patterns with the same
parameters of Fig. 1 but with a larger spread of the initial
wavepacket, $\overline{\sigma}=4 \mu m^{-1}$. Now, the situation
drastically changes with patterns clearly different for bosons and
distinguishable particles, and again for fermions.
\begin{figure}[H]
\center
\includegraphics[width=8cm,height=6cm]{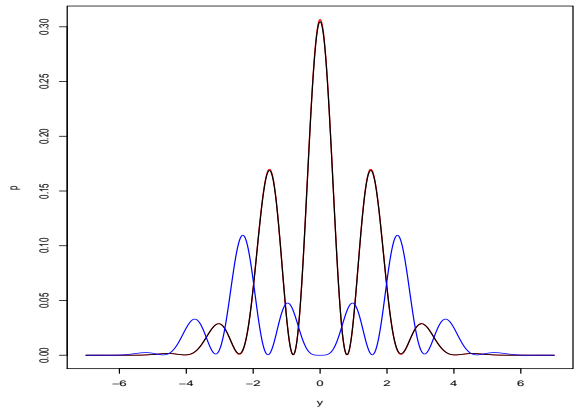}
\caption{Probability density of simultaneous detection (in $\mu
m^{-2}$) versus position of the moving detector $y$ (in $\mu m$) for
the values signaled in the text. The black, red and blue curves
correspond respectively to distinguishable particles, fermions and
bosons.}
\end{figure}

\begin{figure}[H]
\center
\includegraphics[width=8cm,height=6cm]{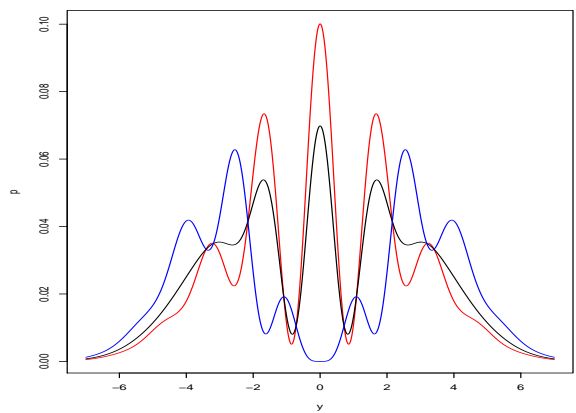}
\caption{The same as Fig. 1 but for $\overline{\sigma}=4 \mu m^{-1}$.}
\end{figure}
The trend observed in the two above figures is general. For all the
$\overline{\sigma}$ smaller than approximately $2.5\mu m^{-1}$ the
curves of distinguishable particles and bosons are almost equal. For
other values of the parameters $b$, $x_0$,... the critical value of
$\overline{\sigma}$ changes, but the behavior is the same. We can
easily understand this property in terms of the final overlapping
between the particles after the diffraction grating. The overlapping
is measured in terms of $|<\psi |\phi>|^2$, which reaches the
maximum value $1$ for full overlapping ($\phi =\psi$) and decreases
to $0$ for orthogonal wave functions with null overlapping. We
represent it in Fig. 3.
\begin{figure}[H]
\center
\includegraphics[width=8cm,height=6cm]{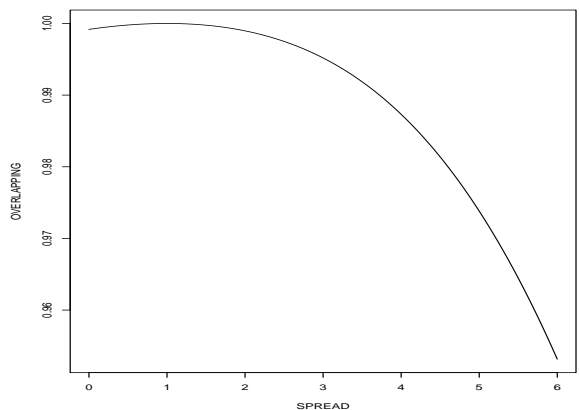}
\caption{Dimensionless overlapping after the diffraction grating
versus initial spread of one of the particles $\overline{\sigma}$
(in $\mu m^{-1}$, with $\sigma$ fixed to $1\mu m^{-1}$).}
\end{figure}
We see that for approximately $\overline{\sigma}<2.5 \mu
m^{-1}$ the final overlapping is very close to unity. For bosons the
full wave function can be expressed as $\Psi_B(x,y) \approx 2\psi
(x)\psi (y)/\sqrt{2+2} \approx \psi (x)\psi (y)$ (because $|<\psi
(x)|\phi (x)>|^2\approx 1$ implies $\psi (x) \approx \phi (x)$). Then
the two-particle wave function for bosons is almost equal to that
of distinguishable particles, a product state, and we must expect
almost equal detection patterns in both cases. For fermions the
situation is different. In the limit of complete overlapping we have,
$\Psi_F(x,y) \rightarrow (1-1)\psi (x)\psi (y)/\sqrt{2-2}$, an
undefined expression of the type $0/0$ reflecting Pauli's exclusion
principle. However, for values very close to unity of the overlapping we
have the quotient of two very small numbers but that is well defined
and, as shown by Fig.1, different from the distinguishable case.

As a byproduct of the above analysis we obtain a potentially
interesting application. We compare the initial and final
overlapping, that is, prior and subsequent to the interaction with
the diffraction grating. The initial overlapping is given by
$2\sigma \overline{\sigma}/(\sigma ^2 + \overline{\sigma}^2)$. We
have that the initial overlapping for $\overline{\sigma}=0.1,0.5,2$
and $4 \mu m^{-1}$ (and $\sigma =1 \mu m^{-1}$) is $0.2,0.48,0.47$
and $0.12$, whereas the final overlapping is $0.99$ for the three
first cases and $0.39$ for the fourth one. The diffraction grating
acts as an efficient mechanism if we use it as a device to increase the
overlapping of the pair of particles.

\section{Experimental tests}

We briefly propose in this section a scheme able to test the
two-atom two-slit patterns with present day technology. We shall
mainly resort to the techniques of atom laser generation. For the
sake of concreteness we shall focus on the arrangement described in
reference \cite{Ess}, which is specially well suited for our
purposes.

Our starting point is the two-particle source. It consists of a trap
containing a large number of atoms (see Fig. 4). We extract atoms
from the trap in a controlled way. This can be done using a
microwave beam. For instance, in \cite{Ess} the $^{87}$Rb atoms in
the magnetic trap are in the hyperfine ground state $|F=1,m_F=-1>$.
The microwave spin-flips some atoms into the state $|F=2,m_F=0>$.
These atoms do not feel the trapping interaction and escape. Then
the atoms fall by gravity and reach the diffraction grating. The
joint patterns are subsequently measured by atom detection within a
temporal coincidence window.
\begin{figure}[H]
\center
\includegraphics[width=8cm,height=6cm]{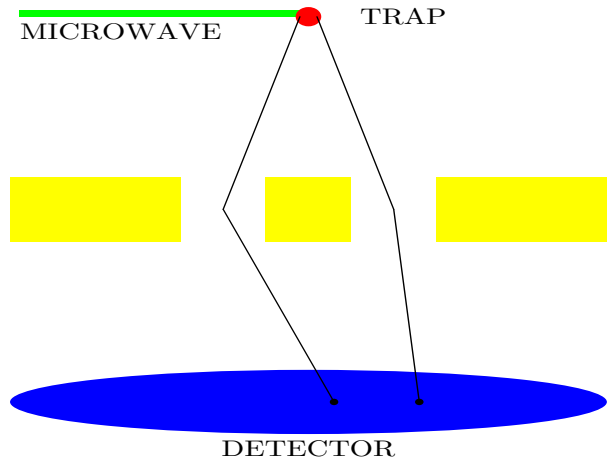}
\caption{The continuous microwave beam ejects atoms from the trap at
a slow rate.}
\end{figure}
The critical points in the scheme are, (i) to make the escaping rate
small enough in order to have small probabilities of a large number
of released atoms in the time interval defined by the coincidence window, and (ii) to guarantee that the atoms detected in that window
have a non-negligible overlapping giving rise to exchange effects.
The two conditions can be fulfilled by the extraction method in
\cite{Ess}. With respect to (ii) it must be remarked that the
microwave beam cannot be monochromatic because this would lead to an
atom laser without bunching. Instead, we must use a broadband beam
with noise (a pseudo-thermal beam), in which bunching effects are
present \cite{Ess}. The presence of these effects certifies the
existence of overlapping between the atoms. Moreover, varying the
chromatic composition of the beam we could control the overlapping
degree. On the other hand, concerning (i), the counting statistics
(for a detection time interval of $1.5 ms$ and a flux close to 5
atoms per $ms$) in the pseudo-thermal case agrees with a Bose
distribution of mean value $1.99$. The probability of time intervals
with five or more detections is small, but the probability of one or
three detections during a single interval is comparable to that of
two events. Then we must introduce a postselection routine to
eliminate the cases with one and three or more detections in the
coincidence window (note, as we shall see later, that the
coincidence window is much shorter than the time interval used to
determine the counting statistics). This routine also eliminates the
cases where two atoms are released from the trap during the
coincidence window but one of them is absorbed by the wall
delimiting the slits.

With respect to the detectors, we suggest to consider the same type
used in \cite{Jel}, which is space- and time-sensitive. This way we
can resolve the point and time of arrival of each particle. The
coincidence window can be close to the values $25 \mu s$ \cite{Jel}
or $50 \mu s$ \cite{Ess} used as time bin sizes to construct the
histograms in these references.

The above discussion has been carried out with bosons. The extension
to fermions follows the same lines. Instead of the BEC of \cite{Ess}
we must consider a trap containing a degenerate Fermi gas.
Using the adequate output coupling we can extract fermions from the
trap in a controlled way.

\section{Discussion}

We have explicitly evaluated the joint detection patterns of the
two-particle two-slit arrangement. The calculations are based on the
extension of Feynman's Gaussian slit approximation to multi-mode
states. This approximation allows for an analytical approach to the problem.

The detection patterns of bosons, fermions and distinguishable
particles are clearly different in the regime of distant mode
distributions (in the space of distribution functions). When these
mode distributions are close the behavior of bosons and
distinguishable particles is almost identical due to the large final
overlapping.

The behavior of bosons in the second case is counterintuitive. With
large overlapping we in general expect the exchange effects to be
also large and the detection patterns of identical and
distinguishable particles to differ. This is so in the HOM
arrangement. In particular, when the photons overlapping is close to
unity (and the reflection and transmission coefficients are equal)
we can observe the famous HOM dip, with no pair of photons leaving
the beam splitter in different arms. This pattern contrasts with that
of no-overlapping photons where the probability for
one photon emerging in each output arm is $1/2$ \cite{Lou}. The
difference between our system and the HOM one lies in the form of
the final states. In our case the almost unity final overlapping
leads to almost equal final wave functions and the approximate form
$2\psi (x)\psi (y)$ for the numerator of $\Psi (x,y)$. The
coefficient $2$, for very large overlapping, cancels with the
denominator, $(2+2|<\psi |\phi>|^2)^{1/2}\approx 2$. $\Psi$
approximately reduces to a product state characteristic of
distinguishable particles. In contrast, in the HOM case the
two-photon state after the beam splitter only adopts
the form corresponding to photons that reflect and transmit
independently (the form we expect for particles that, in the absence of exchange interactions, evolve independently) when the overlapping is null: for large overlapping there is not a cancelation effect of the type present in our arrangement. Up to our knowledge, this counterintuitive type of behavior of the bosons has not been previously described in the literature.

The differentiated behavior of the three types of particles shows
the simultaneous presence of interference and exchange effects. In
particular, we have a manifestation of exchange interactions
potentially verifiable and different from (anti)bunching. This is
interesting because (anti)bunching tests have dominated the
experimental arena in the search of effects associated with the
statistics of massive particles during the last years (see
\cite{Jel} and references therein).

We have also shown that interference at the diffraction grating is
an efficient method to generate large degrees of overlapping. This
point could be of interest in the preparation of (anti)symmetrized
states.

A scheme to experimentally test the results of this paper in atomic
systems has been presented. It is based on existent techniques and
then seems to be a realistic proposal. It can be easily extended to
other types of two-particle interferences and could, with minor
modifications, be used to test the arrangements discussed in
\cite{Sil,SaJP,SaPR}. In the same line of this proposal we could
also test simultaneously interference and exchange effects in
many-particle systems instead of only two-particle ones. By
similitude with the experiment in \cite{Jel} we can turn off the
trapping potential, releasing all the atoms at the same time. After
interaction with a diffraction grating and detection, the observed
correlation functions would contain information on the interference
device parameters and particle statistics. The many-particle
experiment would be much less demanding than the two-particle one.

I acknowledge support from Spanish Ministerio
de Ciencia e Innovaci\'on through the research project
FIS2009-09522.

\end{document}